\documentclass{kapproc} % Computer Modern font calls

\usepackage{t1enc}

\usepackage{procps}

\usepackage[dvips]{graphicx}

\upperandlowercase

\kluwerbib % will produce this kind of bibliography entry:

\newcommand{\IZ}{IZw~36}

\newcommand{\IZW}{IZw~18}

\newcommand{\SBS}{SBS~0335-052}

\newcommand{\Mark}{Markarian~59}

\newcommand{\1}{\,{\sc i}}

\newcommand{\2}{\,{\sc ii}}

\usepackage{epsf}

\begin{document}

\articletitle{Is the interstellar gas of starburst galaxies well mixed?}

%\articlesubtitle{This is an Article Subtitle}

\author{Vianney Lebouteiller}

\affil{Institut d'Astrophysique de Paris, CNRS, 98 bis Boulevard Arago,
F-75014 Paris, France}

\email{leboutei@iap.fr}

\author{Daniel Kunth}

\affil{Institut d'Astrophysique de Paris, CNRS, 98 bis Boulevard Arago,
F-75014 Paris, France} \email{kunth@iap.fr}

\begin{abstract}

The extent to which the ISM in galaxies is well mixed is not yet
settled. Measured metal abundances in the diffuse neutral gas of
star--forming gas--rich dwarf galaxies are deficient with respect to
that of the ionized gas. The reasons, if real, are not clear and need to
be based on firm grounds. Far-UV spectroscopy of giant H\2\ regions such
as NGC604 in the spiral galaxy M33 using \textit{FUSE} allows us to
investigate possible systematic errors in the metallicity derivation. We
still find underabundances of nitrogen, oxygen, argon, and iron in the
neutral phase by a factor of~$\sim$6. This could either be explained by
the presence of less chemically evolved gas pockets in the sightlines or
by dense clouds out of which H\2\ regions form. Those could be more
metallic than the diffuse medium.
\end{abstract}

\begin{keywords}
ISM:abundances - galaxies:starburst - galaxies:ISM - HII regions
\end{keywords}

\section*{Introduction}

The fate of metals released by massive stars in H\2\ regions where stars
are forming is not yet settled. Kunth \&\ Sargent (1986) have suggested
that the H\2\ gas can
% of the dwarf blue compact galaxy (BCD) \IZW\
enrich itself with metals expelled by supernov\ae\ and stellar winds
during the timescale of a starburst. However ionized regions in the LMC
and SMC present very little dispersion in their metal content while
X-ray studies  show that metals reach the halo of galaxies in a hot
phase before they could cool down and finally mix within the ISM in a
few 10$^9$yr. While H\2\ regions abundances derived from optical
emission-lines are usually believed to be representative of the
metallicity of extragalactic regions, the derived abundances would not
necessarily reflect the actual abundances of the ISM if H\2\ regions are
self-polluted.

%An approach to study possible self-enrichment of H\2\ regions and more
%generally the mixing of heavy elements in the interstellar medium (ISM)
%is to compare the H\2\ gas composition derived from optical
%emission-lines with the surrounding diffuse neutral gas derived from
%absorption-lines with \textit{FUSE}.

Blue compact dwarf galaxies (BCDs) are prime targets for the study of
their neutral gas using far-UV absorption lines. These galaxies are
thought to be chemically unevolved and the outskirts of their neutral
cloud could still be pristine. The fate of the newly--produced  metals
in these objects is not clear. A possibility is that, once released by
massive stars they remain in a hot phase, being unobservable immediately
through optical and UV emission lines (Tenorio-Tagle~1996). On the other
hand, Kunth \& Sargent (1986) suggested that heavy elements released by
supernov\ae\ lead to a prompt self--enrichment of H\2\ regions in the
timescale of a star--formation burst. This is supported by Recchi et al.
(2001) model by which most of the newly--synthetized metals mix within
the cold gas phase in a few 10$^6$~years (Myr). Tenorio-Tagle (1996) and
Recchi~et~al. (2001) models differ in the delay between the release and
the final mixing, which can reach several 10$^9$~years (Gyr) in the
first case but only several 10$^6$~years (Myr) in the second.

 A real surprise came from recent \textit{FUSE} study of five BCDs, \IZW\
(Lecavelier et al. 2004 and Aloisi et al. 2003), \Mark\ (Thuan et al.
2002), \IZ\ (Lebouteiller et al. 2004), and \SBS\ (Thuan et al.
submitted). In these objects, nitrogen is systematically underabundant
in the neutral phase as compared with nitrogen abundances derived from
the ionized gas (see Fig.1). Oxygen is either identical in both the
ionized and neutral phases or deficient in the H\1\ gas. The overall
picture brings a new view into the chemical evolution of the ISM in a
galaxy. However it is possible that these results suffer from many
uncertainties such as ionization corrections, depletion effects or
systematic errors due to both multiple sightlines and multiple H\2\
regions within the slit.

In this context, nearby giant H\2\ regions provide a much simpler case
since only one region falls into the aperture, reducing possible
systematic errors. The study of NGC604 presented here is part of a
larger project involving several nearby giant H\2\ regions.

%The
%\textit{FUSE} observations using LWRS and MDRS apertures allow us to
%determine for the first time the neutral gas chemical composition while
%the HST/\textit{STIS} spectrum gives the possibility to map the neutral
%gas inhomogeneities and investigate possible multiple line of sight
%effects.

\begin{figure}[h!]

 \epsfxsize=4.0cm
\rotatebox{-90}{\epsfbox{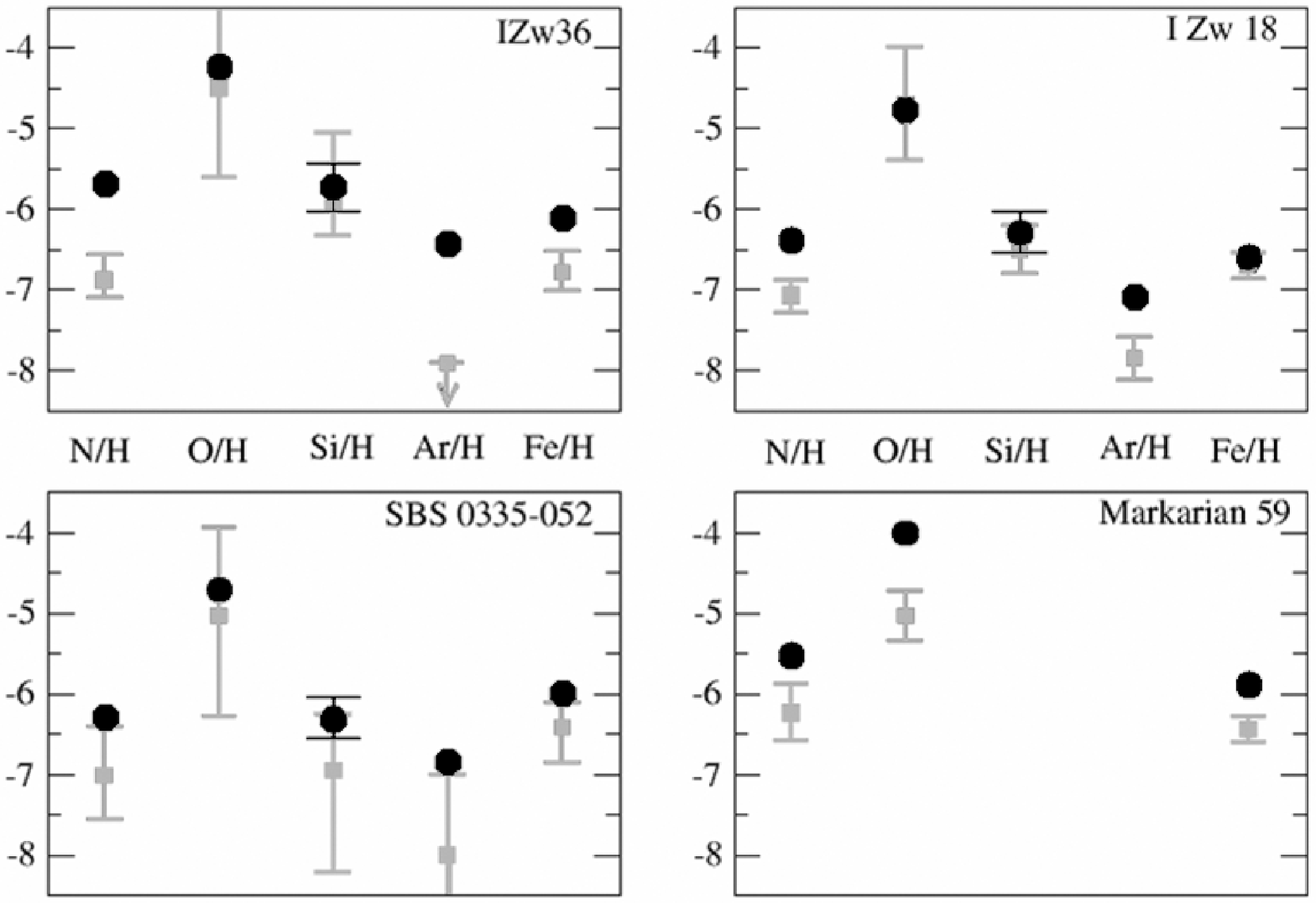}} \hspace{0.3cm} \epsfxsize=4.0cm
 \rotatebox{-90}{\epsfbox{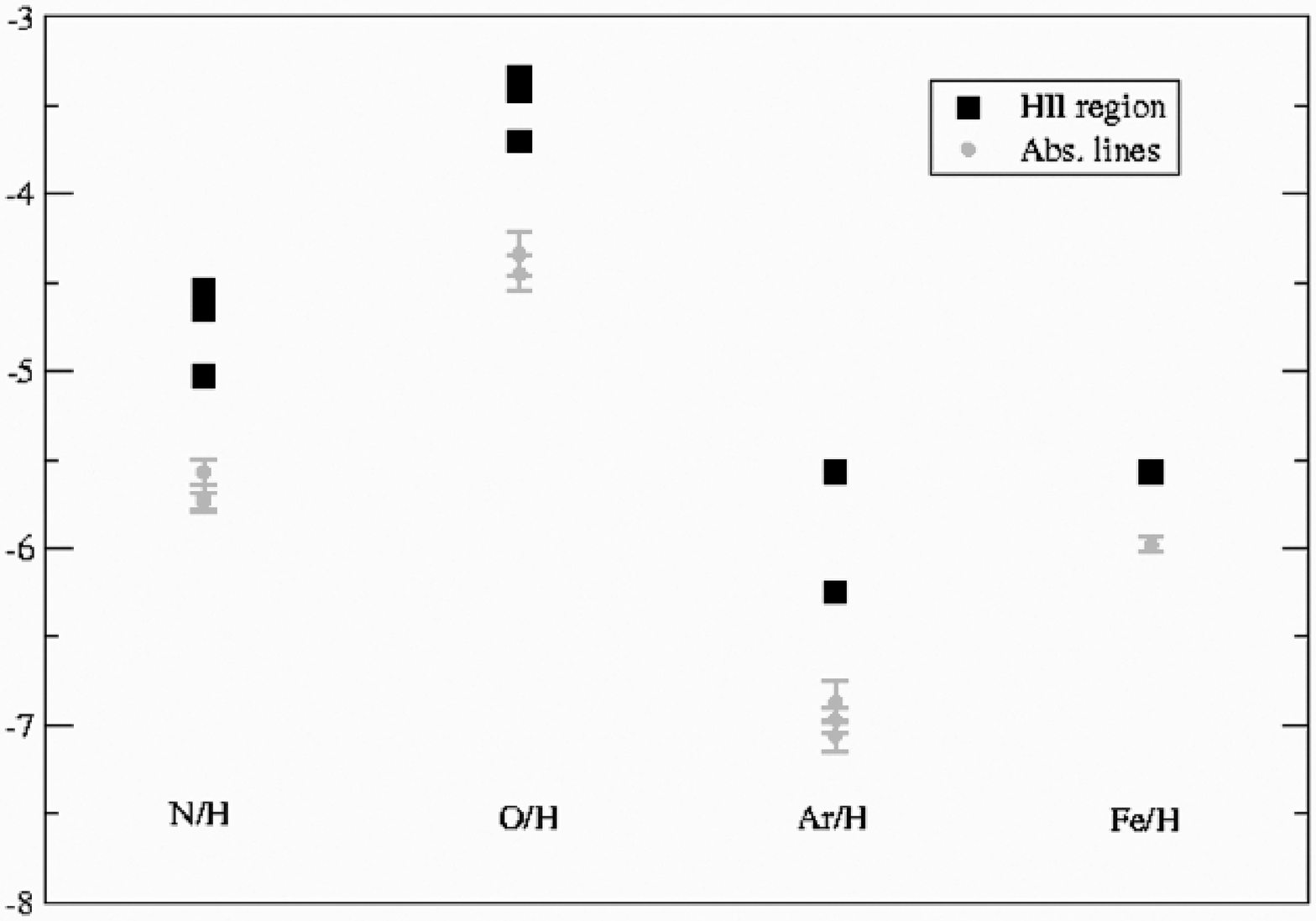}}
 \caption[]{\small{Comparison of log(X/H) in
BCDs (left) and in NGC604 (right) between the neutral gas  abundances
derived from absorption-lines with \textit{FUSE} (in grey) with the H\2\
gas abundances derived from optical emission-lines (in black). Errorbars
at 2$\sigma$. References for ionized gas values: Izotov et al. (1997)
for BCDs, Esteban et al. (2002) for NGC604. References for neutral gas
 values: Lecavelier et al.
(2004) for IZw18, Lebouteiller et al. (2004) for IZw36, Thuan et al.
(2004, in preparation) for SBS0335-052, Thuan et al. (2002) for
Markarian 59 and Lebouteiller et al. (2004, in preparation) for
NGC604.}}

\end{figure}

\section{The case of NGC604}

 The \textit{FUSE}
observations (LWRS and MDRS apertures) allow us to determine the
chemical composition while the HST/\textit{STIS} spectrum gives the
possibility to map the neutral gas inhomogeneities and investigate
possible multiple line of sight effects.

\subsection{Data analysis}

Data analysis has been performed using the profile fitting procedure
\texttt{Owens} developed at the Institut d'Astrophysique de Paris by
Martin Lemoine and the \textit{FUSE} French Team. This program returns
most likely values of many free parameters such as heliocentric
velocities, turbulent velocities, or column densities by a $\chi^2$
minimization of absorption lines profiles. Errors on parameters include
uncertainties on all the free parameters, in particular the position and
shape of the continuum.

We checked the shape of the adopted continuum by comparing the observed
spectrum with a theoretical model of young stellar populations. No
significant difference is found between the model and the continuum we
adopted for the profile fitting. Moreover, by comparing the data with
the model, we find no significant contamination of neutral interstellar
lines  by stellar atmospheres.

By investigating the two \textit{FUSE} observations, we show that an
additional broadening of the lines can account for the spatial
distribution of the bright sources within the slit. The column densities
we derive account for this extension. For the first time, we could check
saturation effects for O\1\ and Fe\2\ lines by analyzing lines
independently. We find no correlation between column densities derived
from each line with the oscillator strength, implying that saturated
O\1\ and Fe\2\ lines in those spectra do not give systematic errors.

\subsection{Neutral gas inhomogeneities}

The high spatial resolution of the HST/\textit{STIS} spectrum of NGC604
gives the possibility to extract spectra towards individual stars of the
ionizing cluster (Bruhweiler et al. 2003).  So far, we have analyzed
three sightlines from which we have measured H\1\ column density using
the Lyman $\alpha$ line. We find spatial variations (up to 0.4 dex)
suggesting inhomogeneities of the diffuse neutral gas. This could be a
source of systematic errors when determining global column densities
from a spectrum of a whole cluster.

We built the global spectrum (i.e. the mean spectrum of all sightlines,
weighted by star magnitudes) of the HST/\textit{STIS} observation to
mimic the spectrum of a cluster in order to compare the real mean column
density we want to determine with the weighted mean we actually measure.
Preliminary results show that we could tend to underestimate the actual
column density when analyzing a global spectrum of several sightlines
towards clouds having different physical properties.

\section{Results and conclusions}

Within errorbars, we derive similar column densities with the two
\textit{FUSE} apertures (see Fig.~1). By modelling the ionization
structure of the H\2\ gas with the photoionization code \texttt{CLOUDY},
we find that N\1\ and O\1\ are good tracers of the neutral gas, contrary
to Ar\1, Fe\2, and Si\2\ which require ionization corrections to obtain
final abundances shown in fig.~1.

We find that N, O, Ar, and Fe are underabundant by approximately the
same factor~$\sim$6 in the neutral phase of NGC604. Whatever the correct
interpretation is, the fact that all specie are equally
 deficient as compared to that of the H\2\
 gas regardless their stellar origin (primary or secondary) is not
 in favor of the self-pollution explanation.
These results could alternatively favor the presence of less chemically
evolved gas pockets in the sightlines, which would tend to  dilute the
metallicity measured in front of the H\2\ region, or that dense clouds
out of which H\2\ regions form could be more metallic than the diffuse
ISM.

%\acknowledgements

%

% This work is based on data obtained for the French Guaranteed

%  Time and the PI Team Guaranteed Time by the NASA-CNES-CSA \textit{FUSE}

%  mission operated by the Johns Hopkins University. French

%  participants are supported by CNES.

%     This work has been done using the profile fitting procedure

%\texttt{Owens.f} developed by M. Lemoine and the

%      \textit{FUSE} French Team.


\begin{chapthebibliography}{1}



\bibitem[Aloisi et al.(2003)]{2003ApJ...595..760A} Aloisi, A., Savaglio,
S., Heckman, T.~M., Hoopes, C.~G., Leitherer, C., \& Sembach, K.~R.\
2003, ApJ, 595, 760

\bibitem[Bruhweiler, Miskey, \& Smith
Neubig(2003)]{2003AJ....125.3082B} Bruhweiler, F.~C., Miskey, C.~L., \&
Smith Neubig, M.\ 2003, AJ, 125, 3082

\bibitem[Esteban, Peimbert, Torres-Peimbert, \&
Rodr{\'{\i}}guez(2002)]{2002ApJ...581..241E} Esteban, C., Peimbert, M.,
Torres-Peimbert, S., \& Rodr{\'{\i}}guez, M.\ 2002, ApJ, 581, 241


\bibitem[Kunth \& Sargent(1986)]{1986ApJ...300..496K} Kunth, D.~\& Sargent,
W.~L.~W.\ 1986, ApJ, 300, 496


\bibitem[Lebouteiller et
al.(2004)]{2004A&A...415...55L} Lebouteiller, V., Kunth, D., Lequeux,
J., Lecavelier des Etangs, A., D{\' e}sert, J.-M., H{\' e}brard, G., \&
Vidal-Madjar, A.\ 2004, A\&A, 415, 55

\bibitem[Lecavelier des Etangs et
al.(2004)]{2004A&A...413..131L} Lecavelier des Etangs, A., D{\' e}sert,
J.-M., Kunth, D., Vidal-Madjar, A., Callejo, G., Ferlet, R., H{\'
e}brard, G., \& Lebouteiller, V.\ 2004, A\&A, 413, 131

\bibitem[Recchi, Matteucci, \& D'Ercole(2001)]{2001MNRAS.322..800R} Recchi,
S., Matteucci, F., \& D'Ercole, A.\ 2001, MNRAS, 322, 800

\bibitem[Tenorio-Tagle (1996)]{} Tenorio-Tagle, G.\ 1996, AJ, 111, 1641

\bibitem[Thuan, Lecavelier des Etangs, \&
Izotov(2002)]{2002ApJ...565..941T} Thuan, T.~X., Lecavelier des Etangs,
A., \& Izotov, Y.~I.\ 2002, ApJ, 565, 941



\end{chapthebibliography}
\end{document}